\documentclass[twocolumn,showpacs,preprintnumbers,amsmath,amssymb]{revtex4}

\usepackage{graphicx}
\usepackage{dcolumn}
\usepackage{bm}

\begin{document}


\title{Rapid non-adiabatic loading in an optical lattice}

\author{Xinxing Liu}
\author{Xiaoji Zhou}\thanks{Electronic address: xjzhou@pku.edu.cn}
\author{Wei Xiong}
\author{Thibault Vogt}
\author{Xuzong Chen}

\affiliation{School of Electronics Engineering and Computer Science,
Peking University, Beijing 100871, People's Republic of China
}%

\date{\today}

\begin{abstract}
We present a scheme for non-adiabatically loading a
Bose-Einstein condensate into the ground state of a one dimensional
optical lattice within a few tens of microseconds typically, i.e. in less than half the Talbot period. This technique of coherent control is based on sequences of pulsed perturbations and experimental results demonstrate its
feasibility and effectiveness. As the loading process is much shorter than the traditional
adiabatic loading timescale, this method may find many applications.
\end{abstract}

.

\maketitle
Numerous works related to optical lattice trapping have been published, especially using
coherent atoms or Bose-Einstein condensate (BEC)~\cite{RMP06}, since it has various applications in quantum computation,
simulation of basic condensed matter physics, atomic clocks, {\it
etc}. A common concern to those experiments is how to load a BEC into the
lattice without excitation or heating~\cite{PRA03,PRA04,PRA05}. In
most cases, one chooses to turn on the light field adiabatically and
the loading process usually lasts up to tens of milliseconds, during which one
tries to avoid excitations to higher states.
A long loading may be problematic for quantum computation experiments as it increases the time during which decoherence can occur and reduces speed when atoms stored in optical lattices shall be interrogated and reloaded several times. Furthermore, adiabatic loading is difficult to obtain near the critical point of phase transitions for finite temperatures even well bellow the BEC critical temperature.

An alternative idea is to prepare the BEC in the ground state of the
lattice, and then turn on the lattice suddenly. This 'preparing'
process can be much shorter compared to the adiabatic loading
(about $30\ \mu s$, as we can see later in this article). One way for realizing this 'preparing' process can stem from nonholonomic coherent control~\cite{PhysRevLett.75.346,PhysRevLett.82.1,EPJ06}. In this approach, a sequence of two well chosen Hamiltonians are imposed on the system and the
preset duration of each step is modified according to a defined cost
function, in order to get the aimed evolution operator as well as
the target state.

Our proposal for designing and computing the pre-loading process is reminiscent of this method, but we focus more on the feasibility of its experimental implementation, rather than the stringency or
universality of the approach. Experimentally, we achieve the non-adiabatic loading with a very few pulses
 using  directly the results of our computational design. According to our
analysis, this method would be valid on many other occasions, such as
loading the condensate directly into excited states of the
lattice.
We first introduce a general method for determining the sequence of steps to be applied to the system before giving an account of our experimental results.

\smallskip

Suppose that before the sudden loading at time $t_0$ of the lattice with optical depth
$V_0$, $m$ steps have been applied. The $i$th step corresponds to a Hamiltonian $\hat{H}_i$ kept constant during $t_i$. The final state $|\psi(t_0)\rangle$ is
given by:
\begin{equation}
\centering|\psi(t_0)\rangle= \prod_{i=m}^1 \hat{U}_i|\psi_0\rangle
\end{equation}
where $\hat{U}_i=e^{-i\hat{H}_i t_i/\hbar}$ is the evolution
operator of the $i$th process. With $|\psi_a\rangle$ the aimed state, which is the ground state
of the lattice in the context of this paper, the total population of the excited
states $N_e$ at time $t_0$, is:

\begin{equation}
\centering N_e=1-|\langle \psi_a|\psi(t_0)\rangle|^2
\end{equation}
Our goal is to
properly choose $\hat{H}_i$ and $t_i$ so that $N_e$ is small enough
to be neglected in the experiment, or the final state $|\psi(t_0)\rangle$ to be close enough to $|\psi_a\rangle$.

The target state $|\psi_a\rangle$ can be obtained simply if atomic interactions are neglected.
The effect of the optical lattice on the atoms corresponds to a stationary
periodical potential with periodicity $\lambda_L/2$, spontaneous emission being neglected since the laser used for the lattice is sufficiently far
detuned. So neglecting atomic interactions, the Hamiltonian of the interaction of atoms with the optical lattice is:
\begin{equation}
\centering\hat{H}_0=\frac{\hat{p}^2}{2m}+\frac{V_0}{2}(1+\cos{2k_Lx})
\label{hamiltonian}
\end{equation}
where $m$ is the atom mass, $k_L$ is the wave number of lattice
light and $V_0$ the lattice depth and $\hat{p}$ is the momentum of each atom. The corresponding Talbot
time~\cite{Deng99} is $T_T=\frac{2\pi}{4\omega_R}$, where
$\omega_R=\frac{\hbar k_L^2}{2m}$ is the one photon recoil
frequency.
Because of the periodicity of the potential, $|\psi_a\rangle$ can be decomposed over a reduced basis of plane waves $|2n\hbar k_L\rangle$. One obvious choice for each $\hat{H}_i$ is to take the Hamiltonian corresponding to the interaction of atoms with a standing wave with same periodicity $\lambda_L/2$ so that the Hilbert space is limited to the basis of states $|2n\hbar k_L\rangle$. For this purpose, the power of the same laser as the one used for the final lattice loading is simply adjusted and each Hamiltonian $\hat{H}_i$ is obtained from Eq. (\ref{hamiltonian}) after substitution of $V_0$ by the new lattice depth $V_i$.

More precisely, as $\hat{H}_i$ has spatial
periodicity, we get its eigenstates by solving the equation
$\hat{H}_i|n,q\rangle=E_{n,q}|n,q\rangle$~\cite{JPB02}, where
$|n,q\rangle$ is a Bloch state with $n$ the band index and $q$ the quasi
momentum and $E_{n,q}$ is the corresponding eigenenergy. We use the notation
$|n,q,V_i\rangle$ for denoting the Bloch states for a $V_i$ lattice depth. Since only Bloch State with $q=0$ is populated initially, no other quasimomenta can be populated during the sequence of pulses. Thus the state of the system can be spanned over the momentum eigenstates basis $|2{\ell}\hbar
k_L\rangle$, independent on the
potential depth $V_i$ and the
evolution operator can be written as the following matrix: $\mathbf
U_i(V_i,t_i)=\mathbf C(V_i)\mathbf{E}(V_i,t_i)\mathbf
C(V_i)^{\dagger}$, where $\mathbf C(V_i)$ is the unitary matrix of transition between the Bloch States basis and the momentum eigenstates basis with matrix elements $\mathbf C(V_i)_{{\ell}n}=\langle 2{\ell}\hbar k_L|n,
q=0,V_i\rangle$ and where $\mathbf E(V_i,t_i)$ is a diagonal matrix with elements
$\mathbf E(V_i,t_i)_{nn}=\exp(-iE_{n,q=0}(V_i)t_i/\hbar)$. Because of the
simple form of the potential, these matrices are easily obtained and from those the wave function's evolution is solved numerically.

A traditional simple approximation to solve this problem of a standing wave of light pulse interaction with atoms is to
omit the effect of kinetic energy. As shown in the
works~\cite{Deng99,Clark10}, the motional term in the Hamiltonian may be
neglected. For a square pulse, when the depth of the standing
wave potential is $V$ and the duration of the pulse is $t$, the wave
function after the pulse can be expressed as $\psi(t)=\psi(0)
e^{-i\frac{V\cdot t}{2\hbar}\cos{2k_Lx}}$ and decomposed over different
momentum states $|2n\hbar k_L\rangle$, using the Bessel functions.
This approximation is valid only in the Raman-Nath regime, when the pulse is
short enough that the displacement of the atoms is much smaller than the
lattice period. From references~\cite{PRL99,NJP09}, we can see experimental
data deviate distinctly from the Bessel functions when the pulse duration
is longer. For multi pulse cases, the motion term may be neglected
during each pulse but taken into account during the
intervals between pulses. This approximation is valid only when the sum of
all the pulses' durations is far less than the Talbot time $T_T$.
In~\cite{Deng99}, $T_T=10\ \mu$s, and two $100\ ns$ pulses were applied,
while the interval time can be up to $10\ \mu$s. However, as the total
pulse duration $200\ ns$ is far less than $T_T$, the whole process is
still in Raman-Nath regime and we can see the theoretical prediction
agrees well with the experimental results.

\begin{figure}
\centering
\includegraphics[height=7cm]{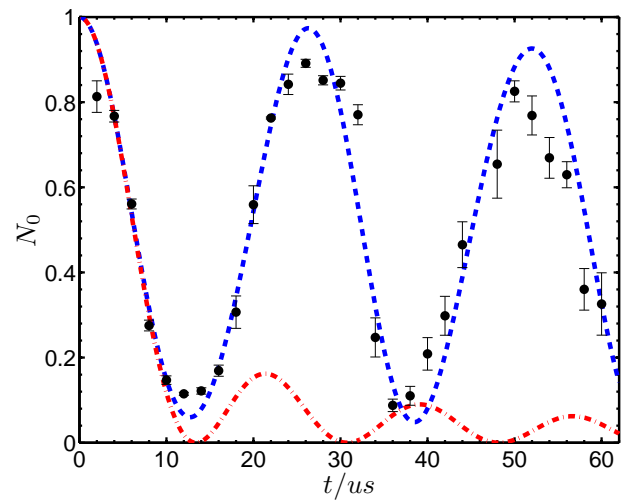}
\caption{(color online) Relative atom number $N_0$ in $p=0\hbar k_L$ mode
 versus the pulse duration $t$ of a single standing-wave pulse applied to the system. The black points with error bars are experimental data and the
red dashed-dotted line is calculated with the simple method where the kinetic energy is neglected
 while the blue dashed line is given by the general method (see text).}
\end{figure}

We compare in Fig.~1 the results of the two methods with our experimental data
for the case of a single square pulse applied to the system. In our experiment, we first prepare a BEC
of about $2\times 10^5$ $^{87}$Rb atoms in the $|F=2,m_F=2\rangle$
hyperfine ground state in the magnetic trap, with longitudinal
length $L=100\ \mu$m and width $l=10\ \mu$m~\cite{Zhou10}. The
BEC is then loaded into a one-dimensional optical lattice with
light wavelength $\lambda_L=852$~nm, along its axial direction. We
suddenly turn on the standing-wave potential to a specific depth,
hold the lattice for different durations $t$, release the condensate and
measure the relative atom number $N_0$ in $p=0\hbar k_L$ mode.
Without considering the atomic motion, the Bessel function (red
dashed-dotted line) failed to fit the experiment data when $t$ is
larger than about $10\ \mu$s. On the other hand, the numerical solution
taking the motional term into account (blue dashed line) predicts
the experiment accurately up to $60\ \mu$s, although some other effects have been neglected, such as interaction between atoms and non-zero
momentum width of the condensate. As a result, we will use the general
method as it will impose less restriction on our sequence design.
Note that $T_T\approx79\ \mu$s in our configuration and the fitted
potential depth is $18E_R$, where $E_R=(\hbar k_L)^2/2m$ is the one
photon recoil energy.

In a first attempt to design the time sequence, we use four steps,
each corresponding to the application of a potential with depth
$V_i$ and a duration $t_i$ as shown in Fig.~2a. Thus, from these
eight free parameters ${V_i, t_i}$ with $i=1$ to 4, we can obtain
the evolution operators ${\mathbf U_i}$ and the exciting rate $N_e$
according to our previous analysis. If we want to get the minimum of
$N_e$, it requires to test all the combinations of the eight free
parameters and need too much computing time. However, we find
convergence to a local minimum is easy to obtain. For this purpose
we modify the parameters step by step, with a small change of one
parameter every step, in order to get a smaller $N_e$. If no smaller
result for $N_e$ can be obtained by changing any of these
parameters, the minimization of $N_e$ is stopped and the
corresponding combination of parameters is the result of our design.
Note that different initial values of the parameters may
result in different local minima. Actually we try
more than one group of initial parameters in order to
get a better result, though the optimized pulse sequences obtained for different and often lower lattice depths are already good input values.
 As shown in Fig.~2c, the exciting rate
$N_e$ remains lower than $0.1\%$. The whole pre-loading process
lasts for about $30\ \mu$s, which is beyond the Raman-Nath Regime
but still within the limit of our method.

\begin{figure}
\centering
\includegraphics[height=9.5cm]{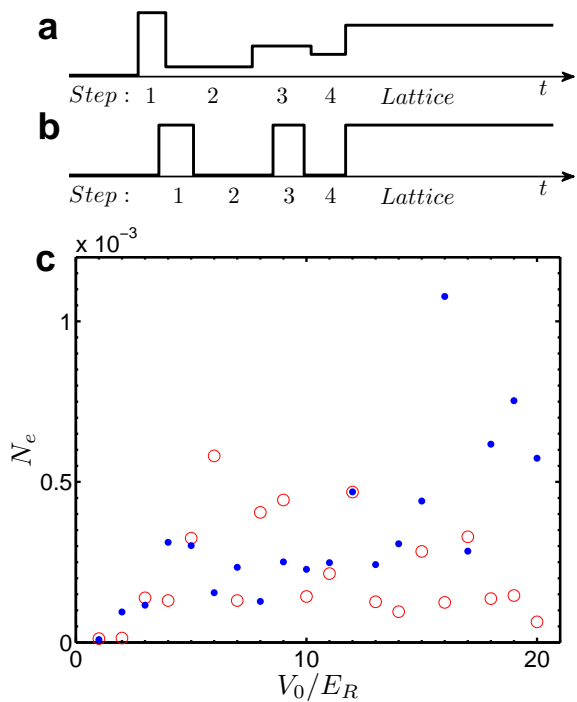}
\caption{(color online) Two schemes for the non-adiabatic
loading. (a) Four steps pre-loading sequence with both the
potential depth and the duration of each step set as free
parameters. (b) Four steps pre-loading sequence with fixed potential
depth and variable duration in each step. (c) Corresponding relative population $N_e$ in excited Bloch
states as a function of lattice depth, with
red circles for scheme (a) and blue points for scheme (b).}
\end{figure}

Nevertheless, it is not easy for this scheme to be experimentally
implemented. As the lattice depth varies with time, we need a feed
back loop to control the laser power to get a stable optical lattice as the atoms
are very sensitive to fluctuation of the laser power. We also need sharp rising and
falling edges to ensure each step duration is sufficiently similar to the
designed one. If $0.5\ \mu$s is set for the upper limit of each rising
and falling edge, the bandwidth of the power controlling loop should
be larger than $1$~MHz, which is hard to achieve. We would rather
choose the scheme illustrated in Fig.~2b, where the step $V_1$ and $V_3$ are
fixed at $V_0$, and the step $V_2$ and $V_4$ are fixed at $0$, $t_i$ still being
free parameters. This means no feed back loop but only a switch is
needed so that the rising and falling time can be decreased to
$200\ ns$ easily. In Fig.~2c, we can see the exciting rate is
still lower than $0.1\%$ for most lattice depths lower than $20\ E_R$,
although some degrees of freedom of our sequence of pulses are locked.
In TABLE I we show the designed time sequences for several
lattice depths.

\begin{table}\caption{Time duration of each step $t_i$ ($\mu$s) for different lattice depths $V_0$ with unit $E_R$.}
\begin{center}
\begin{tabular}{ l || c c c c | c c c r  }
\hline\hline
$V_0$ & $t_1$ & $t_2$ & $t_3$ & $t_4$&$t_5$ & $t_6$ & $t_7$ & $t_8$ \\
\hline 4&10 &9.5 &4  &5 &6  &5 &9   &9
\\  8&6.5 &10.5  &5  &6 &6   &4.5 &10.5  &7
\\  12&5   &11.5    &4.5 &6  &5.5 &4   &11.5  &5.5
\\  16&4.5 &12  &3.5 &6 &5   &4   &12.5   &4.5
\\  20&3.5 &13  &3.5 &5.5 &6   &3.5 &12.5  &3.5
\\
\hline\hline
\end{tabular}\end{center}
\end{table}

Furthermore, we can transfer the state $|\psi_a\rangle$ back to
the initial state $|\psi_0\rangle$ after releasing the lattice, using a similar process.
TABLE I gives also the designed sequence, with time durations $t_5$ to $t_8$ and corresponding potentials, $V_5$ and $V_7$ being set to $0$, while $V_6$ and $V_8$ equal
to $V_0$ (see Fig.~3b). Note that this sequence is equivalent to the timely inverted pre-loading sequence if we ignore some minor
deviations.

\begin{figure}
\centering
\includegraphics[height=12cm,width=9cm]{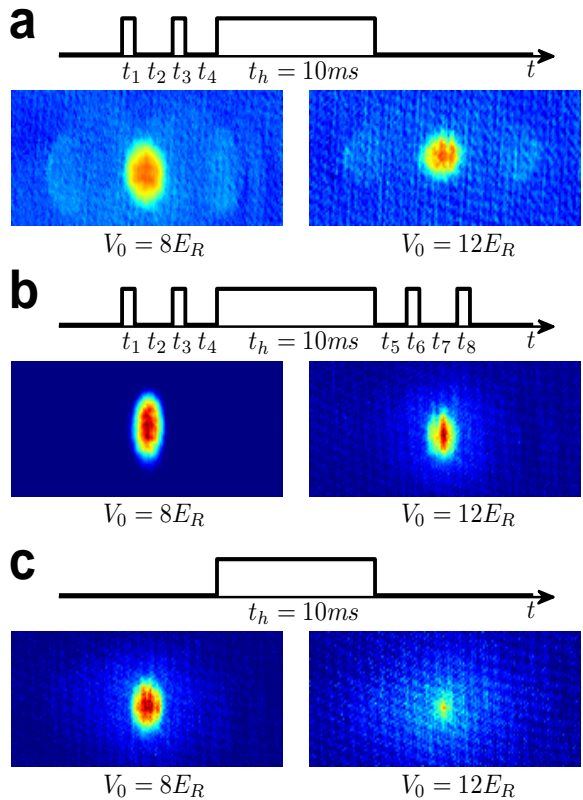}
\caption{(color online) Absorption images for three different
time sequences. As the condensate was released from both the lattice
potential and the magnetic trap, it expands freely for $30ms$ and
then an absorption image is taken. For each time sequence, we show the results for two
lattice depths of $8E_R$ and $12E_R$.}
\end{figure}

To demonstrate the feasibility of our proposal, we did the
experiment according to three different schemes. In the first scheme
(Fig.~3a), we non-adiabatically load the BEC into the lattice, according
to the designed process, hold it in the lattice for $t_h=10\ ms$ and release the atoms. We can see the interference peaks, similar to the familiar pattern observed in adiabatic loading experiments, which
indicates a successful loading without significant excitation and
heating. In the
second scheme (Fig.~3b), after the same loading and holding process,
we use two additional pulses to transfer the atoms back to the
original state $|\psi_0\rangle$. Compared with the third scheme
(Fig.~3c), where we directly turn on and off the lattice light
without the pre-loading or post-releasing processes to prevent
excitation but hold for the same period, the second scheme has
little heating or disturbing effect on our BEC, which proves the
effectiveness of our 'preparing' process of the lattice ground state. A small heating may
still be observed at $12E_R$. It may be due to interaction between atoms which are not included in our model.

\begin{figure}
\centering
\includegraphics[height=7cm,width=8cm]{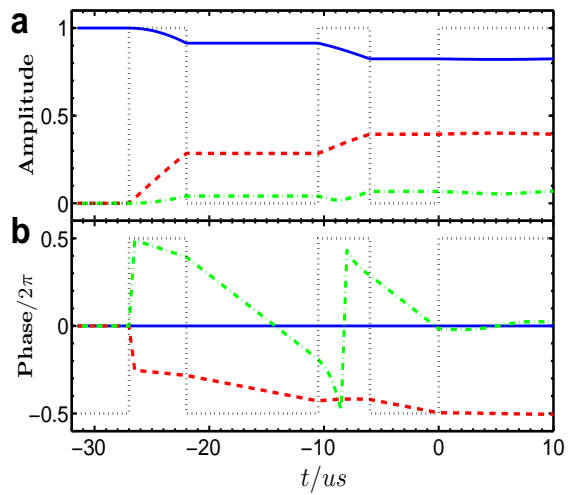}
\caption{(color online) Evolution of the different momentum components of the wave function. Amplitude (a) and phase (b) for the $p=0\hbar k_L$, $\pm 2\hbar k_L$ and $\pm
4\hbar k_L$ modes are shown with blue solid lines, red dashed
lines and green dashed-dotted lines respectively. The black dotted
lines in both figures indicate the corresponding time
sequence of the standing wave potential.}
\end{figure}

Fig.~4 shows the evolution of the wave function under a pulsed
scheme shown in Fig.2b. We can see that atoms can be transferred between different
momentum orders only when the pulse is on. In the intervals, the
amplitudes of different components keep constant, while the phases
vary linearly. The phase of the $\pm 4\hbar k_L$ order evolves four times
faster than that of $\pm 2\hbar k_L$ as the kinetic energy is four
times lager. After switching on the optical lattice ($t=0\ \mu$s), the amplitude and phase of all
components keep almost unchanged as the wave function is in the ground state of the lattice. Note that the
phase jump of the non-zero momentum components at $t=-27\ \mu$s is not
significant since the amplitude of these components before that
time is zero.

By simply replacing the initial state $|\psi_0\rangle$ and the aimed
state $|\psi_a\rangle$, our pulse sequence design is also applicable to other situations. For example, by setting $|\psi_0\rangle=|n=0, q=0,
V_1\rangle$ and $|\psi_a\rangle=|n=0, q=0, V_2\rangle$, we can get a
process that can change the lattice depth from $V_1$ to $V_2$
non-adiabatically without excitation. We can also load the BEC directly
to symmetric excited states such as $|n=2, q=0\rangle$, by setting
them as $|\psi_a\rangle$. Our numerical simulation shows when $V_0$ is
around $12\ E_R$, there exists theoretical possibility to load more
than $99\%$ of the atoms to $|n=2,q=0\rangle$. This state can be used for studying spontaneous transition from high lying Bloch bands to lower
bands experimentally. If the optical lattice is accelerated, anti-symmetric states and states with $q\neq0$ could also be loaded.

In general, our method can be used to produce states whose momentum
components are discrete and separated equally by $2\hbar k_L$. For
example, we can divide one BEC into two momentum modes $|\pm2m\hbar
k_L\rangle$ equally, with negligible population in other orders.
This technique is useful in atom interferometry and its theory has been developed
in reference~\cite{Clark10}, while neglecting the motional term in the
Hamiltonian, as discussed earlier in this article. Our method is not
restricted by the Raman-Nath regime and thus gives more freedom for designing the pulse sequence.

In conclusion, we proposed a method reminiscent of the nonholonomic
coherent control technique for loading the BEC into one dimensional
optical lattice non-adiabatically, within a much shorter loading
time (usually less than $T_T/2$) than the commonly applied
adiabatic method (much longer than $T_T$). Our experimental results
demonstrate the validity of this method. We claim that this numerical design
process can be applied to various topics related to the interaction of a standing wave of light with ultra-cold gases.

This work is supported by NKBRSFC(2011CB921501) and
NSFC(61027016,61078026,10874008 and 10934010).

\end{document}